\documentclass[aps,twocolumn,prl,superscriptaddress,amsmath,showpacs,tightenlines]{revtex4}
\usepackage{graphicx}
\usepackage{epsfig}

\begin{document}

\title{Scalability of spin FPGA: A Reconfigurable Architecture based on spin MOSFET}

\author{Tetsufumi~Tanamoto, 
        Hideyuki~Sugiyama, 
        Tomoaki~Inokuchi,
        Takao~Marukame,
        Mizue~Ishikawa,
        Kazutaka~Ikegami,
        and~Yoshiaki~Saito}

\affiliation{Advanced LSI Technology Laboratory
	Corporate Research and Development Center,
	Toshiba Corporation
	1, Komukai Toshiba-cho, Saiwai-ku,
	Kawasaki 212-8582, Japan.}

\date{\today}

\begin{abstract}
Scalability of Field Programmable Gate Array (FPGA) using spin MOSFET (spin FPGA) 
with magnetocurrent (MC) ratio in the range of 100\% to 1000\% is discussed 
for the first time. Area and speed of million-gate spin FPGA are numerically 
benchmarked with CMOS FPGA for 22nm, 32nm and 45nm technologies 
including 20\% transistor size variation. We show that area is reduced and speed is 
increased in spin FPGA owing to the nonvolatile memory function of spin MOSFET.
\end{abstract}

\maketitle

\section{Introduction}
Spin metal-oxide-semiconductor field-effect transistor (spin MOSFET) is a novel  MOSFET 
whose source and drain are contacted with ferromagnetic materials~\cite{Tanaka}. 
Ferromagnetic materials provide stable and robust nonvolatile memory~\cite{Ohno}. 
Fig.1(a) shows a spin MOSFET in which the write process is carried out by using magnetic tunneling junction (MTJ)~\cite{Marukame,Inokuchi}. 
Spin MOSFET directly couples logic element with nonvolatile memory element, 
opening up a path to a new style of logic-in-memory architecture~\cite{Kautz}. 

Field Programmable Gate Array (FPGA) has a great advantage in that a chip is 
completely programmable and reconfigurable.
However, conventional FPGA includes a lot of static random access memory (SRAM), 
which is a volatile memory composed of six transistors and faces the fabrication limitation
 of Si MOSFET. Thus, new FPGA based on novel devices has been expected.
Here, for the first time, we report on numerical benchmark for an island-style FPGA using 
 22nm, 32nm and 45nm spin MOSFETs (spin FPGA)~\cite{Inokuchi}
by improving standard benchmark tools~\cite{Betz}.
Compared with other proposals\cite{DeHon,Dong}, spin FPGA has an advantage in that it is 
based on Si transistor equipping stable nonvolatile magnetic memory.
Moreover, SRAM (six transistors) can be replaced by one spin MOSFET.
Many SRAMs are used in FPGA such as in Lookup tables~(LUTs) and 
interconnect area of pass transistors. 
Therefore, this replacement reduces transistors and FPGA area. 
Because the speed of FPGA is governed by the length of wire part,
smaller area of spin FPGA leads to faster  performance. 
Monte Carlo simulation based on the Predictive Technology Model~\cite{PTM} 
is carried out to consider variation of device size assuming fabrication difficulties.
Although experiments on MTJ~\cite{Ohno} at present 
show the maximum magnetocurrent (MC) ratio is 260\% 
($RA\approx 10 \Omega \mu$m${}^2$), in this paper 
we treat $100\% \le{\rm MC \ ratio} \le 1000\% $  assuming future realization of larger MC.

\section{Spin FPGA}
{\it Spin MOSFET.}---We model the spin MOSFET by changing SPICE parameter 
(mobility) such that MC 
defined by ${\rm MC}=(I_{\rm P}-I_{\rm AP})/I_{\rm AP}$ coincide with a given MC ratio
($I_{\rm P}$ and $I_{\rm AP}$ are parallel and antiparallel currents, respectively.) 
For $I_{\rm P}$, we use the same SPICE parameters as those of the conventional MOSFET (Fig.1(b)).
Although there is extra resistance owing to the existence of MTJ in spin MOSFET, 
as Ref.\cite{Nagamine} reported, the resistance of 50~nm square MTJ can be controlled to less than 400$\Omega$
and this resistance is negligible compared to the resistance of conventional MOSFET 
of the order of 10~k$\Omega$. 

\begin{figure}[!t]
\centering
\includegraphics[width=8.5cm]{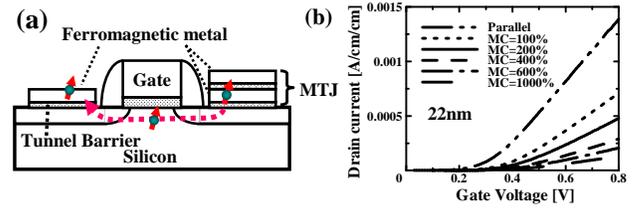}
\vspace*{2.5cm}
\caption{(a) Spin-based MOSFET in the type of ``Spin-transfer-Torque-Switching MOSFET''
in which magnetic tunnel junction (MTJ) are attached to one of the electrodes.
(b) $I_d$-$V_g$ characteristics for parallel and antiparallel states ($100\% \le {\rm MC} \le 1000\%$) based on PTM SPICE model (see text).}
\label{fig_spinMOS}
\end{figure}

\begin{figure}[!t]
\centering
\includegraphics[width=2.5in]{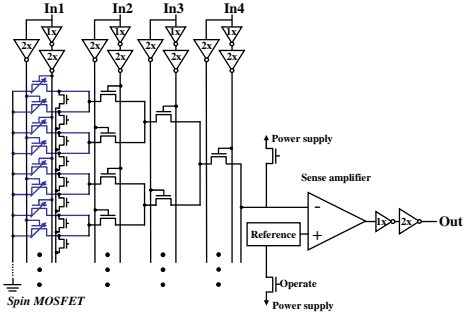}
\vspace*{4cm}
\caption{Schematic of a 4-input look up table based on spin MOSFET (spin LUT).
Spin MOSFETs replaces SRAMs at the leftmost part of this figure.}
\label{spinLUT}
\end{figure}

\begin{figure}[t]
\centering
\includegraphics[width=8.5cm]{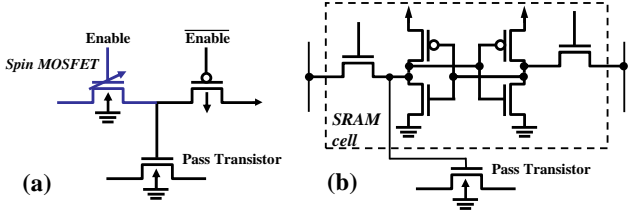}
\vspace*{3.0cm}
\caption{(a) New routing pass transistor using spin MOSFET and (b) that using conventional SRAM.
In (a), one extra transistor is required to change P/AP state of  
spin MOSFET, and the width of spin MOSFET and PMOS is enlarged to
control ON/OFF state of attached pass transistor. The estimated number of 
required control transistors in (a) is four in minimum-width transistor area model~\cite{Betz}.}
\label{fig_pass}
\end{figure}

{\it Spin Cluster Logic Block.}---Fig. \ref{spinLUT} shows our spin LUT structure~\cite{Sugiyama} for 4-inputs and 1-output, 
which is a typical set of LUT parameters~\cite{Betz}. 
Transistor sizes of amplifiers are adjusted such that the input pulse signal 
is appropriately transferred to the output of LUT.

{\it Pass transistor.}---
We propose a spin control pass transistor depicted in Fig.~\ref{fig_pass} (a). 
SPICE simulations show that the speed of pass transistor in Fig.~\ref{fig_pass}(a) is 
of the same order as that in Fig.~\ref{fig_pass}(b) by adjusting the width of control transistors
(total transistor area of Fig.\ref{fig_pass}(a) is four in unit of minimum transistor size). 
Although this pass transistor structure has a disadvantage, namely, a leakage pass from $p$-type transistor (PMOS) 
to $n$-type transistors (NMOS),
this power dissipation can be reduced by limiting the on-state  
only when it is required~\cite{Lundstrum}.

%%%%%%%%%%%%%%%%%%%%%%%%%%%%%%%%%%%%%%%%%%%%%%
\section{FPGA area reduction by spin MOSFET}
First, let us compare the number of transistors in spin LUT and CMOS LUT.
In ref.~\cite{Sugiyama}, we only counted the number of transistor of a spin LUT. 
Here, we estimate the number of transistors by a general clustered logic block (CLB) in which 
four CLBs are clustered with 10 inputs and 4 outputs.
For $K$-input  LUT, $2^K$ SRAM and $2^{K+1}-2$ pass transistors (multiplexer trees) are required 
with three input buffers. 
Then the total number of transistors in a 
complementary MOS (CMOS) LUT $N_{\rm lut}^{\rm (cmos)}$ is given by $2^{K+3}-2+6K$. 
In a spin LUT (Fig.\ref{spinLUT}),  the leftmost SRAMs are replaced by spin MOSFETs 
with an additional write/erase transistor. 
In addition, a sense amplifier (five transistors), a reference transistor and two power supply transistors 
are required. Thus, the number of transistor required in the spin LUT is 
given by $N_{\rm lut}^{\rm (spin)}= 3\times 2^K+6(K+1)$.
Thus, we have  $N_{\rm lut}^{\rm (cmos)}-N_{\rm lut}^{\rm (spin)}=5 \times 2^K-8$. 
For example, 
4-input LUT conventionally has 150 transistors whereas spin LUT includes 78 transistors (48\% reduction).

Circuit area is calculated by the minimum-width transistor area model~\cite{Betz}, 
in which each transistor area is estimated by a unit of minimum-width NMOS.
When $W_{\rm min}$ and $S_{\rm min}$ are width and area of minimum NMOS, respectively,  
a width $Z W_{\rm min}$ transistor is estimated as having an area of $(1+Z)S_{\rm min}/2$.
%===========  pmos =================
Width of PMOS is determined such that an inverter changes at 
half of  a drain voltage. For PMOSs of 22nm, 32nm and 45nm nodes, 
\begin{equation}
Z_{\rm 22nm}^{\rm (pmos)}=1.53, \ Z_{\rm 32nm}^{\rm (pmos)}=2.22, \ Z_{\rm 45nm}^{\rm (pmos)}=2.57
\label{ZZ}
\end{equation}
(PMOS is scaled down more than NMOS because of advanced technologies such as strain effects.)
Area of recent FPGA is mostly occupied by an interconnect or wiring part.
Wire resistance and capacitance are calculated from Ref.~\cite{ITRS}.

\section{Benchmark Results and discussion}
\begin{figure}
\centering
\includegraphics[width=7cm]{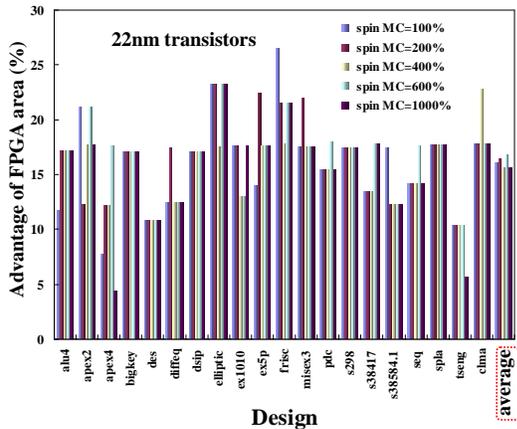}
\vspace*{5.3cm}
\caption{Benchmark calculation of the advantage of spin FPGA to CMOS FPGA over 
20 circuits (area). Rightmost data shows average over the 20 circuits.}
\label{area22}
\end{figure}

\begin{figure}
\centering
\includegraphics[width=7cm]{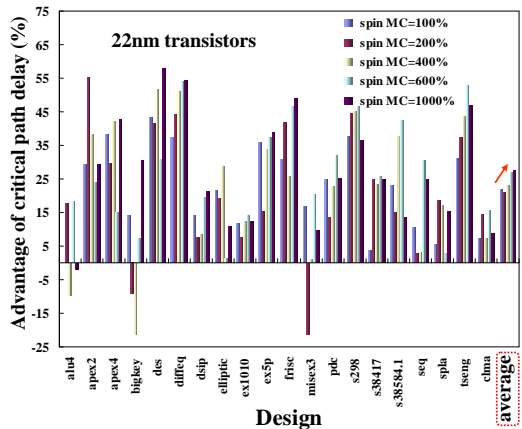}
\vspace*{5.3cm}
\caption{Benchmark calculation of the advantage of spin FPGA to CMOS FPGA over 
20 circuits (delay). Mean critical path delay of CMOS FPGA is 30.8 ns and 
those of spin FPGA are 25.4 ns (MC=100\%), 25.5ns (MC=200\%), 25.2 ns (MC=400\%), 24.5ns (MC=600\%) and 24.5ns (MC=1000\%). 
}
\label{delay22}
\end{figure}

\begin{figure}
\centering
\includegraphics[width=7cm]{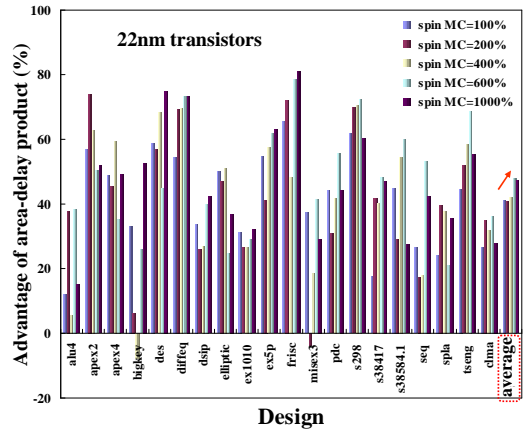}
\vspace*{5.4cm}
\caption{Benchmark calculation of the advantage of spin FPGA to CMOS FPGA over 
20 circuits (area-delay product). 
}
\label{AD22}
\end{figure}

\begin{figure}
\centering
\includegraphics[width=5cm]{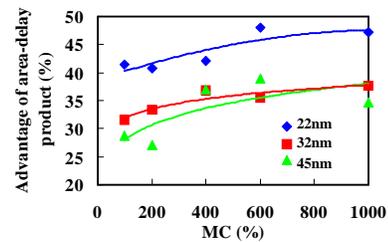}
\vspace*{2.8cm}
\caption{Comparison of transistor generation. 
An average result of the benchmark calculation as a function of MC ratio.
Relations between generations are considered to be related with relative PMOS areas (see Eq.(1) and text).}
\label{scaling1}
\end{figure}

Area and speed of spin FPGA over 20 typical million-gate circuits are benchmarked with modified VPR ver.5~\cite{Betz}
for 22nm, 32nm and 45nm transistors. 
We take standard parameters such as $F_s=3$ (Wilton switch box), $F_c\_{\rm in} =1.0$ and 
$F_c\_{\rm out} =0.25$  with length 1 wire segment~\cite{Betz}.
Fig.\ref{area22}-\ref{AD22} show the average results over 200 Monte Carlo simulations 
for up to 20\% (3 sigma) variations of length and width in 22~nm transistors, 
where the vertical axes show  advantage of area, critical path delay and area-delay product
defined by $(\Theta^{\rm cmos}-\Theta^{\rm spin})
/\Theta^{\rm spin}$ for $\Theta$=\{\rm $A$ (area), $t_{\rm delay}$ (critical path delay), 
$A\times t_{\rm delay}$ (area-delay product)\}.
Area-delay product is treated as a metric of FPGA performance.
Fig.\ref{area22} and Table I show that area of spin FPGA is greatly reduced 
compared with CMOS FPGA. For 22~nm transistor, an average of 16\% 
area reduction is realized. This area reduction leads to small critical path delay of circuits 
resulting in faster operation in spin FPGA.
In Fig.~\ref{delay22} speed is improved by an average of 24\%.
As MC ratio increases, P/AP signals that go into an amplifier in spin LUT (Fig.~\ref{spinLUT}) 
become clearer. This leads to more robust operation against the variation of transistors, 
resulting in shorter delay in Fig.~\ref{delay22}.
Thus, area-delay product is improved on average by 43\%. 
Fig~\ref{scaling1} shows summarized results of benchmark from 22~nm to 45~nm transistors.
As mentioned above, as transistor scale decreases, ratio of PMOS area to NMOS area 
decreases. This means that the effect of area reduction by spin MOSFET (NMOS) 
becomes larger resulting in better performance of small transistor nodes.

%\small
\begin{table}
\begin{tabular}{c||c|c||c|cccc}
\hline 
            & \multicolumn{2}{c||}{CLB area ($\mu$m${}^2$) }    & \multicolumn{5}{c}{ Interconnect area ($\times 10^{3}$)($\mu$m${}^2$)}   \\ \cline{2-8}
            &     CMOS &SpinMOS   &   CMOS              & \multicolumn{4}{c}{ SpinMOS }  \\ 
            &      &   &        & 100\%   & 200\%              &600\%               & 1000\% \\ \hline
22nm    & 118.6   & 97.2    & 237.8  & 207.0 & 208.2 & 206.9 & 208.1 \\ 
32nm    & 124.1   & 102.7    & 242.8  & 210.3 & 211.6 & 207.8 & 212.1 \\ 
45nm    & 250.7   &  208.3   & 483.0  & 416.7 & 418.4 & 421.8  & 413.9 \\ 
      \hline 
\end{tabular}
\caption{Area of  a single CLB and interconnect. Result of interconnect is taken from Fig.~\ref{area22}.}
\label{segment}
\end{table}

One of the advantages of spin MOSFET compared with CMOS with interlayer MRAM system 
is that, for spin MOSFET, MC ratio change directly affects subthreshold region of MOSFET 
which leads to more efficient device operations. The effect 
of direct injection of spin into channel on device performance 
will be clarified in more detail in the near future.

\section{Conclusion}
Spin FPGA was numerically benchmarked for 22nm, 32nm and 45nm transistors. We showed 
that the performance of spin FPGA becomes superior to that of conventional CMOS FPGA 
as transistor size decreases and MC ratio increases.

\end{document}